\documentclass[a4paper,aps,pre,twocolumn,groupedaddress,showkeys,showpacs,floats,floatats,floatfix]{revtex4-1} 
\usepackage[utf8]{inputenc}
\usepackage[english,brazil]{babel}
\usepackage{graphicx}
\usepackage{dcolumn} 
\usepackage{bm}
\usepackage{natbib}
\usepackage{latexsym}
\usepackage{mathrsfs}
\usepackage{amssymb}
\usepackage{amsmath}
\usepackage{amscd}
\usepackage{color}
\usepackage{pifont}
\usepackage[usenames,dvipsnames]{xcolor}
\usepackage{pstricks,pst-node,pst-text,pst-3d}
\usepackage{verbatim}
\usepackage{ulem}
\newrgbcolor{Green}{0.00 0.432 0.078}
\newrgbcolor{Red}{0.7 0.1 0.1}
\newrgbcolor{Blue}{0.2 0.2 0.6}
\newrgbcolor{Cyan}{0.0 0.8 0.8}
\newrgbcolor{Magenta}{1.0 0.0 1.0}
\newrgbcolor{Orchid}{0.6 0.2 0.8}
\newrgbcolor{Orange}{0.9 0.6 0.0}
\newrgbcolor{Gold}{0.62 0.42 0.0}
\newrgbcolor{Pink}{1.0 0.08 0.58}

\begin{document}
%

\title{Recurrence-time statistics in non-Hamiltonian volume preserving maps and flows}
\author{Rafael M.~da Silva$^{1,2}$}
\email{rmarques@fisica.ufpr.br} 
\author{Marcus W.~Beims$^{1,2}$}
\email{mbeims@fisica.ufpr.br}
\author{Cesar Manchein$^{1}$}
\email{cesar.manchein@udesc.br}

\affiliation{$^1$Departamento de F\'\i sica, Universidade do Estado 
de Santa Catarina, 89219-710 Joinville, Brazil}
\affiliation{$^2$Departamento de F\'isica, Universidade Federal do Paran\'a, 
Caixa Postal 19044, 81531-980 Curitiba, Brazil}
\date{\today}
%
\begin{abstract}
We analyze the recurrence-time statistics (RTS) in  three-dimensional
non-Hamiltonian volume preserving systems (VPS): an 
extended standard map, and a fluid model. The extended map is 
{a standard map} weakly coupled to an extra-dimension which 
contains a deterministic regular, mixed (regular and chaotic) 
or chaotic motion. The extra-dimension strongly enhances the trapping 
times inducing plateaus
and distinct algebraic and exponential decays in the RTS plots. The
combined analysis of the RTS with the classification of ordered and
chaotic regimes and scaling properties, allows us to describe the
intricate way trajectories penetrate the before impenetrable regular
islands from the uncoupled case. Essentially the plateaus found
in the RTS are related to trajectories that stay long times inside 
trapping tubes, not allowing recurrences,
and then penetrates diffusively the islands (from the uncoupled case) by 
a diffusive motion along such tubes in the extra-dimension. All 
asymptotic exponential decays for the RTS are related to an ordered
regime (quasi-regular motion) and a mixing dynamics is conjectured
for the model. These results are compared to the RTS of the standard
map with dissipation or noise, showing the peculiarities obtained by
using three-dimensional VPS. We also analyze the RTS for a fluid model and 
show remarkable similarities to the RTS in the extended standard map 
problem.  
\end{abstract}



\maketitle
\section{Introduction}
\label{sec:introduction}
The seminal work of Henri Poincar\'e brought up to attention a very
important fact: the motion of an incompressible fluid of finite volume
is recurrent \cite{poincare1890}. For Hamiltonian systems this means
that almost all orbits come arbitrarily close to an initial point an
infinite number of times. For many years the properties of such
recurrences have been analyzed in distinct physical situations by
using the RTS. It was observed that the RTS is capable of describing
the relevant aspects of the dynamics in complex systems. In this
context, we mention that the RTS is able to describe universal
algebraic decays in Hamiltonian systems
\cite{altmann06,shepelyansky10,AltmannThesis},  
including random walk penetration of the Kolmogorov-Arnold-Moser (KAM) 
islands~\cite{eduardoPRL10,bernal13}, biased random walk to escape from 
KAM island \cite{ketzmerick12}, DNA sequence \cite{shepelyansky}, synchronization of 
oscillator \cite{afraimovich00}, generalized bifurcation diagram of conservative
systems \cite{beims13}, fine structure of resonance islands \cite{egydio13}, 
transient chaos in systems with leaks \cite{altmann09}, among others.

In distinction to the above works related to area-preserving maps, 
a novel situation occurs for non-Hamiltonian three-dimensional 
($3$D) VPS, whose dynamics has applications in volume-preserving flows,
and chaotic scattering, where particles are captured and scattered
on resonance (see Ref.~\cite{mezic-chaos} for a detailed discussion). 
One work in this direction related the RTS with anomalous transport in a 
fluid flow \cite{zas91}. {In case of scattering problems, there exist a 
scattering region where interactions between particles and the system  
of interest occur. Outside this region the action of the 
system is insignificant so that the particles motion is not of relevance. 
Essential properties from the system of interest is revealed from the behavior 
of the scattered particles. In fact, scattering is one of the fundamental tools 
to unveil the main characteristics of processes in nature, such as in chemical 
reactions, atomic and nuclear physics, fluid and celestial mechanics 
\cite{Lichtenberg,telbook,sanju13}. Recent attentions focused on scattering in 
conservative systems, where regular and chaotic dynamics coexist and
particles can be trapped for finite times close to regular islands
leading to an algebraic temporal decay of the scattered particles,
which perform a transient chaotic motion. In realistic experiments,
the scattered particles may also be subjected to noise, dissipation,
external forces, etc. In most cases noise destroys the KAM curves and
the algebraic decay is replaced by the exponential
\cite{grassberger84,ottbook,tel00}, or stretched  exponential
\cite{grigolini14}. However, recent developments have shown that
noise can enhance the algebraic decay due to trajectories performing a
random walk motion inside the KAM
islands~\cite{eduardoPRL10,bernal13}. Biased random walk was   
also observed in the standard map under random symplectic
perturbations \cite{ketzmerick12}. 

On the other hand, flows in three-dimensions arise in a number of different 
applied contexts such as fluid dynamics and transport of matter in a fluid 
\cite{Khurana,mezic03,torney07}. Nonlinear maps can be obtained from fluid
flow models and are of two types: action-action-angle and action-angle-angle 
maps. Whereas a action-angle-angle volume-preserving map is analogous to 
symplectic maps when it comes to KAM theorem, the behavior of 
action-action-angle maps is very different. It was shown \cite{PiroFeingold} 
that no invariant $2$D surfaces persist upon perturbation from an integrable 
action-action-angle map and at these same locations in the phase-space periodic 
orbits of specific type persist and they dominate the transport in the perturbed
map. Besides, many initial conditions can create trajectories that go through a 
large portion of the phase-space even when the map is very close to being 
integrable. This has been named {\it resonance-induced dispersion} (RID)
\cite{passivescalars,CartwrightFeingoldPiro,msm1}. 

In this work we analyzed the RTS for two systems: First an
action-action-angle map, called the Extended Standard Map ({\it ESM}),
introduced in Sec.~\ref{sec:model}, a non-Hamiltonian $3$D 
VPS which can be associated with the chaotic scattering problem 
mentioned above. Our discussion will be mainly focused in the
({\it ESM}) system. Secondly we consider a fluid flow model \cite{mezic03} 
to show the astonishing similarities to the RTS in the  {\it ESM}.
For the case of the 3D non-Hamiltonian map considered (the ESM),
we} use an additional deterministic dimension, weakly coupled to the 
standard map (vortices in the fluid case), which allows particles to
penetrate the regular islands from the uncoupled case, and eventually
approach the center of the island. Mainly we show that the
extra-dimension strongly enhances the trapping time around the regular
island from the standard map (vortices), resulting in plateaus 
(not allowing recurrences) and algebraic decays in the characteristics
recurrence curves and generating asymptotic exponential decays as
typically observed, for instance, in area-preserving maps.
While short time exponential and long time algebraic decays are 
similar to those found in Hamiltonian system, the large plateaus are
apparently a particular property of $3$D VPS. We mention here that the
plateaus were also found in a totally distinct context, a fertility
model which has some applications to earthquakes \cite{saichev13}. In
addition, asymptotic exponential  decays appear to be common in $3$D 
VPS. In our analysis we combine the RTS and a recent developed technique 
\cite{marques15}, which uses finite-time Lyapunov exponents (FTLEs), to 
efficiently determine the time evolution of ordered (${\bf O}$), and chaotic 
(${\bf C}$) regimes. This allows us to clearly associate the decays of 
cumulative distribution of consecutive time spent in the regime ${\bf O}$ 
(algebraic or exponential), with the penetration inside the regular 
islands from the uncoupled case. Scalings properties of the RTS show that 
the physical origin of the asymptotic decays is the same for different small 
couplings intensities with the extra-dimension.

Finally, with this work 
we also intend to give some insight about possible answer for the
following questions: ``Do recurrence-time distributions in typical 
volume-preserving maps with regular and chaotic components have an 
asymptotic power-law form like Hamiltonian systems with
mixed dynamics? Is there an universal decay exponent of 
recurrences? Similar questions were proposed recently \cite{meiss14}, 
where the author summarizes both, the state of the art in the theory 
of transport for conservative dynamical systems, based 
on the last thirty years of investigations, and tries to point out some
open problems that could be addressed next.

This work is organized in the following manner. While Sec.~\ref{sec:model} 
presents the three-dimensional volume preserving model used to describe
our results, Sec.~\ref{sec:methods} briefly describes the methods of the RTS and 
classification of regimes using the time dependent finite-time Lyapunov spectrum. 
In Sec.~\ref{sec:results} the results for the ESM are discussed, and
are compared to the standard map with dissipation or noise in
Sec.~\ref{uncoupled}. Section~\ref{flow} discusses the results for the
fluid flow model and concluding remarks are summarized in
Sec.~\ref{sec:conclusion}. 

\section{The coupled maps model}
\label{sec:model}
Our main benchmark tool in this work is the {\it ESM} defined by
\renewcommand{\arraystretch}{2.2}
\begin{equation}
\label{msm}
\left\{
\begin{array}{lllllll}
p_{n+1}=p_{n}+\dfrac{K_1}{2\pi} \sin(2\pi x_{n})\hspace{0.2cm}+&\dfrac{\delta}{2\pi} 
\sin(2\pi w_{n})&\mathrm{mod} \hspace{0.1cm} 1, \\
w_{n+1}=w_{n}+\dfrac{K_2}{2\pi} \sin(2\pi x_{n}) & \mathrm{mod} 
\hspace{0.1cm} 1,\\
x_{n+1}=x_{n}+p_{n+1}&\mathrm{mod} \hspace{0.1cm} 1,\\
\end{array}
\right.
\end{equation}
\renewcommand{\arraystretch}{1}
\noindent
where the parameter $K_1$ determines the dynamics of the
area-preserving standard map, the parameter $K_2$ establishes the
dynamics of the extra-dimension as regular, mixed or chaotic, and 
the parameter $\delta$ is a measure of the intensity of the coupling
between the standard map and the extra-dimension. Some analytic 
properties of this model are investigated in Ref.~\cite{msm1}. 

For $\delta=0$, {the} map {given by Eq.}~(\ref{msm}) reduces to the
usual standard map in $x$ and $p$ coordinates, while $w$ behaves
like another action coordinate \cite{msm2}. The standard map has a
well known rich  dynamics \cite{Lichtenberg, Reichl04}, which changes 
from regular for small $K_1$, to mixed for intermediate values of
$K_1$ or to totally chaotic for large $K_1$. For clarity, in 
Fig.~\ref{phase-space-mp} we plot the phase-space of the standard map
for the uncoupled case, showing the (a) regular dynamics, (b) and (c)
the mixed dynamics, and (d) the chaotic dynamics. The coupled dynamics
for $\delta > 0$ intermixes the motions observed in
Fig.~\ref{phase-space-mp} and the resulting regular structure in $3$D
phase-space can have a complicated shape.
To be precise with the terminology, for the uncoupled case we 
will use {\it regular island} (which cannot be penetrated), while for the 
coupled case we use {\it regular structure} (which can be penetrated).
\begin{widetext}
$\quad$
\begin{figure}[!htb]
  \centering
  \includegraphics*[width=1\columnwidth]{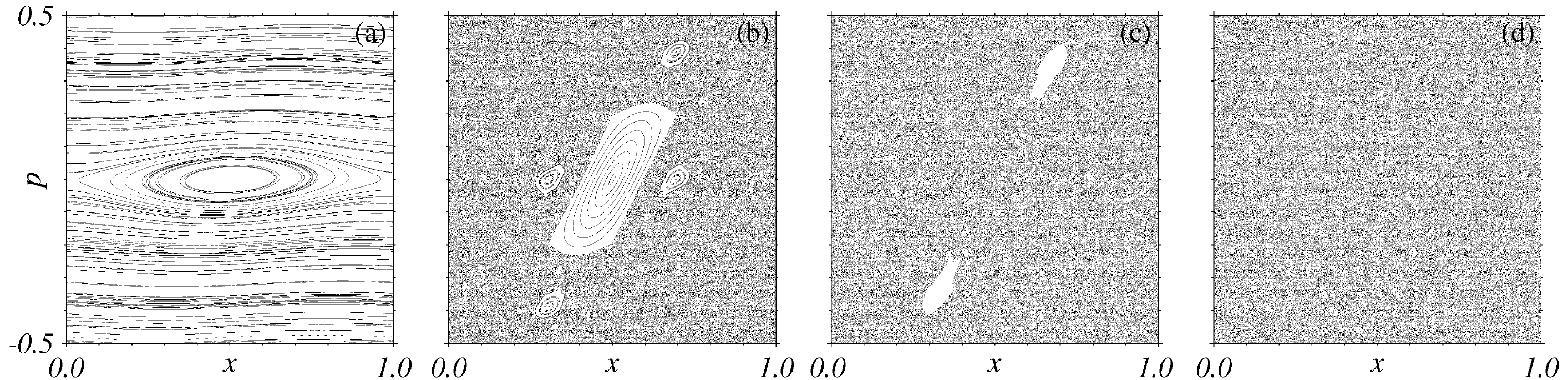}
  \caption{Phase-space dynamics of the uncoupled ($\delta=0$) standard map 
  using (a) $K_1=0.09$ (regular dynamics), (b) $K_1=2.6$
  (mixed dynamics), (c) $K_1=4.9$ (mixed dynamics) and (d)
  $K_1=8.8$ (chaotic dynamics). These values will define the dynamics of 
  the extra-dimension coupled to the standard map (see Sec.~\ref{sec:results}). 
  Here we used $80$ randomly {equally distributed} ICs and $3\times10^4$
  iterations for each IC.}
  \label{phase-space-mp}
\end{figure}
\end{widetext}

A stability analysis, not presented here, shows that the stable 
period-$1$ and $4$ resonances shown in Fig. \ref{phase-space-mp} become
unstable for $\delta > 0$. Therefore, the center of the regular structure 
becomes unstable and has crucial consequences for our results. There might 
exist other regular structures with stable centers for this model, but we 
have not found them. In fact, other kind of couplings and/or extra-dimension 
dynamics could 
be used to keep the stability of the center of the structure, but this will
certainly change the results presented in Sec.\ref{sec:results}.

\section{Methods}
\label{sec:methods}
Two methods, or techniques, will be used to carefully describe the influence 
of the extra-dimension on the scattered trajectories, as briefly
  discussed in the Sec.~\ref{sec:introduction}. 

\subsection{Recurrence-time statistics}
The first technique is the RTS determined numerically by computing the
quantity of iterations $\tau$ that the trajectory, starting from an
initial pre-delimited chaotic region, collides with the
regular structure and comes back to the initial region,
called recurrence region.  We are interested in the cumulative
probability distribution $P_{cum}(\tau)$ defined by 
\begin{equation}
P_{cum}(\tau) \equiv \displaystyle \sum_{\tau'=\tau}^{\infty} P(\tau'), 
\label{Pcum}
\end{equation}
\noindent
that gives essential informations about the presence, or not, of sticky motion 
in conservative systems. Straight lines in the log-log plot of this distribution 
are power-law decays of the form $P_{cum}(\tau)\propto \tau^{-\gamma}$, where 
$\gamma$ is the decay exponent. Such decays are usually related to sticky effects 
induced by the invariant structures on the chaotic trajectory
\cite{AltmannThesis}. To reckon $P_{cum}(\tau)$ we always start with
ICs randomly distributed inside the  
recurrence region defined by $-0.2\le x \le0.2$, $\-0.5\le p \le0.5$
and $-0.5 \le \omega \le 0.5$. Each time the recurrence region is 
mentioned, it means that the trajectory is already ejected from the 
region close to the regular structure. We also tested other
boxes for the recurrence region, with different sizes, but our main
results remain unchanged.   

\subsection{Lyapunov Regimes}

\begin{figure}[!htb]
  \centering
  \includegraphics*[width=0.95\columnwidth]{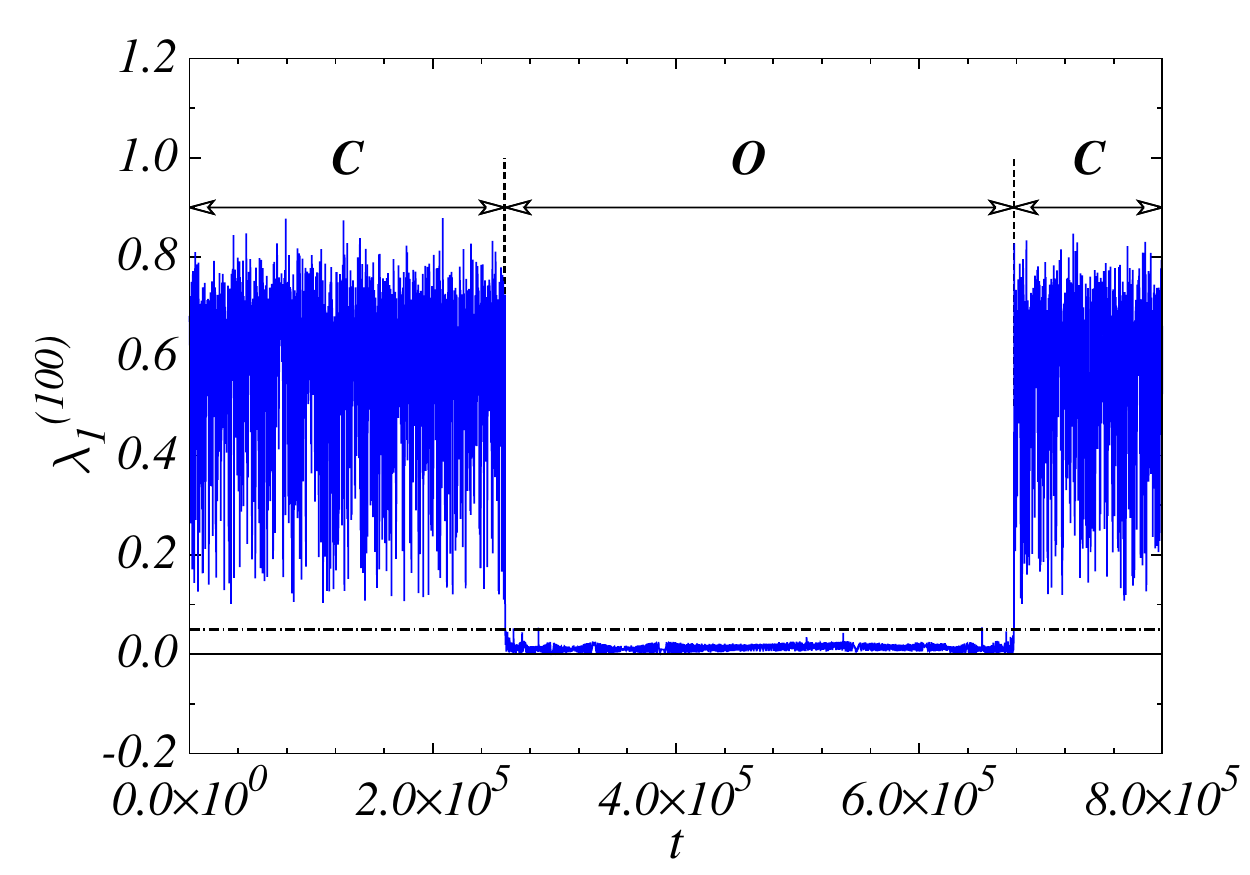}
  \caption{(Color online) Illustration of the method used to
  define the regimes $\bf O$ and $\bf C$. The time series of the 
  FTLE $\lambda_1^{(\Omega=100)}$ for the {\it ESM} is plotted. 
  The threshold $\varepsilon=0.05$ that defines the regimes is plotted by a 
  dashed-dotted line. A typical initial condition used to  generate this time 
  series was chosen inside the chaotic sea.}
  \label{lyaptme}
\end{figure}

The second technique consists in using the FTLEs spectrum
$\{\lambda_i^{(\Omega)}\}$ to classify ordered and chaotic regimes in  
time as proposed recently \cite{marques15}. For the {\it ESM} model we
have three FTLEs $(i=1,2,3)$ which satisfy 
$\lambda_3^{(\Omega)}=-\lambda_1^{(\Omega)}$ and $\lambda_2^{(\Omega)}=0$.
In order to implement this technique, we follow a trajectory and
compute the FTLEs spectrum during a window of size $\Omega$, and
explore the temporal properties in the time series of  
$\lambda_1^{(\Omega)}$. Starting with a typical initial condition
chosen in a chaotic region where the largest FTLE is greater than zero,
we follow the trajectory and for times $t=m\Omega$ ($m=1,2,\ldots,N$)
we evaluate $\lambda_1^{(\Omega)}$. If this trajectory is close to the regular
structure we have $\lambda_1^{(\Omega)} \approx 0.0$. This allows us
to classify distinct regimes as {\it ordered} ${\bf O}$, for
$\lambda^{(\Omega)}_{1} \approx 0$, and {\it chaotic} ${\bf C}$, for
$\lambda^{(\Omega)}_{1} > 0$, as exemplified in Fig.~\ref{lyaptme}
along a time series of $\lambda^{(100)}_{1}$. In the regime ${\bf O}$
the trajectory is close to the regular structure and 
for regime ${\bf C}$ it is in the chaotic region of the phase space.
From this analysis we can obtain the cumulative probability 
distribution defined as
\begin{equation}
  P_{cum}(\tau_M) \equiv \sum_{\tau'_M=\tau_M}^{\infty} P(\tau'_M), 
  \label{PcuM}
\end{equation}
where $\tau_M$ is the time spent consecutively in one of the two possible regimes 
and $M= {\bf O},{\bf C}$  indicates the corresponding regime. It is worth to mention 
that the ordered regime is composed by a chaotic trajectory which moves, for a 
finite-time, close 
to a regular island (sticky motion). It has been shown \cite{marques15} that 
this classification of regimes significantly improves the ability of the 
stickiness characterization and, by associating them with the distinct decays 
in the RTS, a clear picture of the underlying dynamics. 

\section{Results}
\label{sec:results}

In this Section the numerical results are
analyzed. For all simulations we use a mixed dynamics ($K_1=2.6$)
for the standard map, and change the dynamics of the extra-dimension
using the following values $K_2=0.09,2.6,4.9$ and $8.8$ (see
Fig.~\ref{phase-space-mp}).  

\subsection{Recurrence-time statistics plots}
\label{RTS}

Figure \ref{rec} (a)-(d) displays the log-log plot of the RTS as a function 
of $\tau$ for different values of $K_2$. For comparison we show the uncoupled 
standard map case $\delta=0$ {(black-dashed line)}, which displays an exponential 
decay for small times followed by the power law decay  ($\sim\tau^{-1.62}$) due 
to sticky effects. For {all} couplings we still have the power-law decay 
($\sim\tau^{-1.62}$ ) {for shorter times.} 
After these times, a plateau is seen whose width depends on the coupling strength.
The plateau in the RTS means that no recurrences were observed for these times. 
As will be shown later, {during the plateaus} the trajectory moves around 
the regular islands from the uncoupled case and do not return to the recurrence 
region. For increasing values of the coupling, we observe that the plateaus appear 
for earlier times and become shorter. This occurs {independently of the dynamics 
in the {extra-dimension}}, but tend to disappear for $\delta\sim 10^{-1}$ (not 
shown here) in the regular case in Fig.~\ref{rec}(a), and for
$\delta\sim 10^{-3}$ {(cyan-dashed-dotted line)} in the mixed and 
chaotic cases shown in Figs.~\ref{rec}(b)-(d). Besides the mixed case from 
Fig.~\ref{rec}(c), which has an abrupt break of the plateau, {where} all plateaus are 
followed by exponential decays for larger times. {In fact, the case 
$\delta\sim 10^{-3}$ in Fig.~\ref{rec}(c) also shows the asymptotic exponential 
decay after the abrupt break of the plateau} and a smooth power law related with the
random walk performed by the trajectory inside the regular
structures. In general we observe the following decays: 
\begin{itemize}
\item $\bf C$: Exponential decay for very short times (not discussed in this work).
\item $\bf O_1$: Power-law with $P_{cum}(\tau)\propto \tau^{-1.62}$ for 
  $\tau<\tau_{p_{\delta}}$, where $\tau_{p_{\delta}}$ is the time when the plateau 
  appears for given $\delta$. This decay is the only one (besides ${\bf C}$) 
  observed for   $\delta=0$, and the first one (after ${\bf C}$) observed for 
  all analyzed couplings. 
\item $\bf O_2$: Plateaus obeying $P_{cum}(\tau)\propto\delta$ for times
  $\tau_{p_{\delta}}<\tau<\tau_{p_{\delta}^{\prime}}$, where 
  $\tau_{p_{\delta}^{\prime}}$ is the time where the
  plateaus disappear.  
\end{itemize}
After the plateaus we see distinct behaviors which depend on the dynamics of the 
extra-dimension:
\begin{itemize}
\item ${\bf O_3}$: An exponential decay with $P_{cum}(\tau)\propto
  e^{-\eta\tau}$ for $\tau>\tau_{p_{\delta}^{\prime}}$. This is the
  asymptotic behavior for the cases $K_2=0.09$ in 
  Figs.~\ref{rec}(a),(e),(i), and $K_2=8.8$ in Figs.~\ref{rec}(d),(h), (l), and
  intermediate decay for $K_2=2.6$ in Figs.~\ref{rec}(b),(f) and (j). 
  
\item ${\bf O_3^{\prime}}$: An asymptotic exponential 
      decay with $P_{cum}(\tau)\propto e^{-\eta\tau}$, only observed in 
      Figs.~\ref{rec}(b),(f) and (j), when using $K_2=2.6$.

\item ${\bf O_4}$: A short power-law with  $P_{cum}(\tau)\propto\tau^{-4.6}$ after the 
      abrupt break of the plateau. This was only observed for $K_2=4.9$ in 
      Figs.~\ref{rec}(c),(g) and (k). Other 
      simulations (not presented here) also show the abrupt break of the 
      plateau for values close to $K_2\sim 4.9$.

\item ${\bf O_5}$: After ${\bf O_4}$ a random walk with $P_{cum}(\tau)\propto\tau^{-0.5}$ 
      occurs (also only for $K_2=4.9$).
\end{itemize}
These distinct behaviors of the RTS in the ordered regime can be 
better visualized in the insets of Figs.~\ref{rec}(a)-(d). Before discussing 
in details the physical origin of the decays, we equip ourselves with 
the regimes technique, introduced in the Sec.~\ref{sec:methods}B,
to have a better understanding of the underline dynamics. 

\begin{widetext}
$\quad$
\begin{figure}[!htb]
  \centering
  \includegraphics*[width=1\columnwidth]{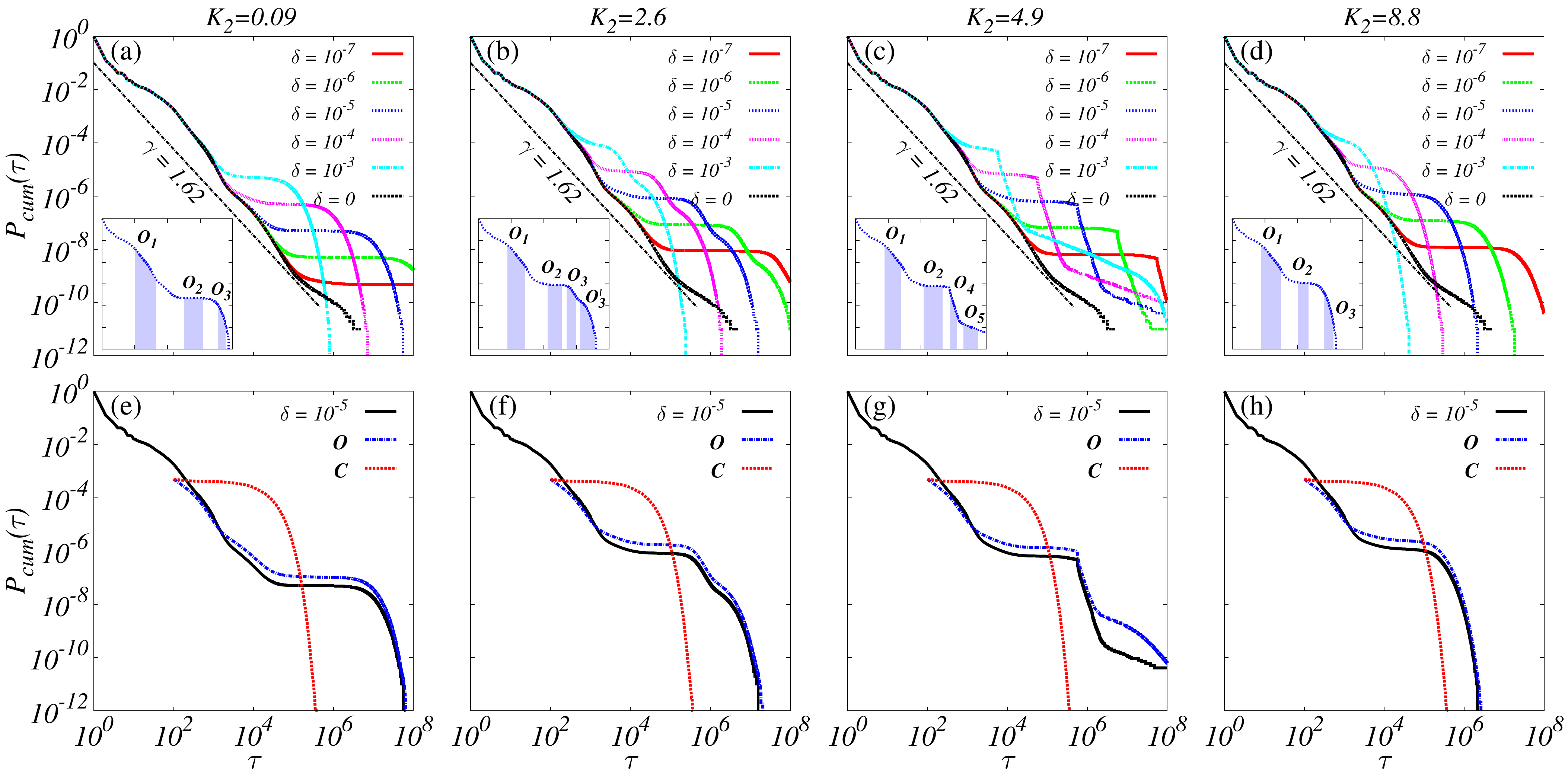}
  \includegraphics*[width=1\columnwidth]{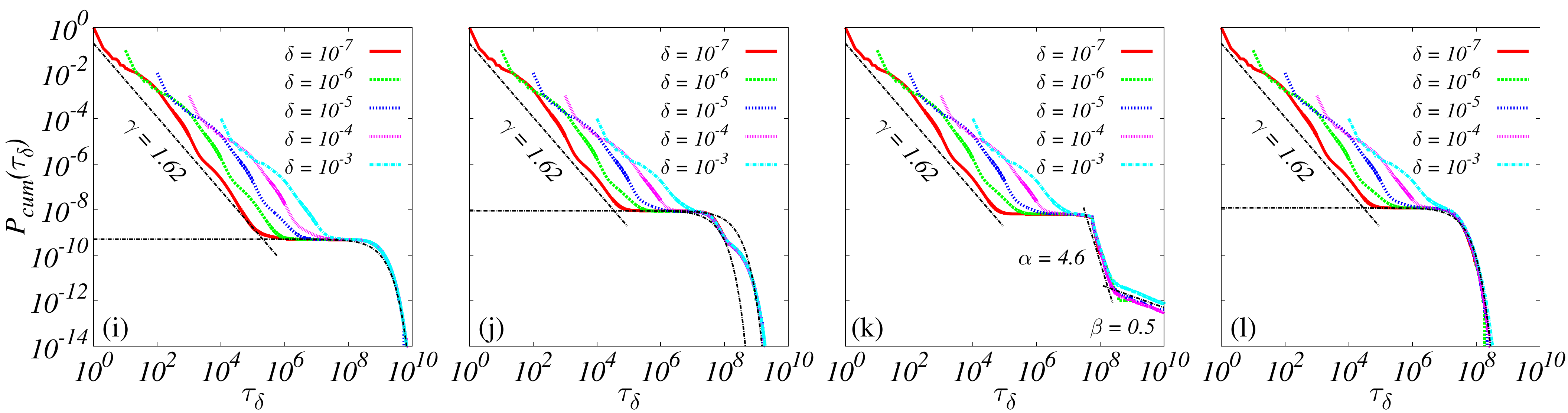}
  \caption{(Color online) In top and bottom lines are plotted the
    cumulative probability distributions $P_{cum}(\tau)$
    [Eq.~(\ref{Pcum})] for recurrence-times $\tau$ and its  scaled
    version $P_{cum}(\tau_{\delta})$ [Eq.~(\ref{scal})] respectively,
    for different values of the coupling $\delta$. For $\delta=0$ 
    we recover results for the standard map. For clarity, the insets in 
    the top line show the distinct behaviors 
    (${\bf O_1}, {\bf O_2}, {\bf O_3}, \ldots$) of
    the RTS in the ordered regime for $\delta=10^{-5}$. In the middle line 
     we also plot $P_{cum}(\tau)$ for recurrence-times $\tau$
     (black-continuous lines) together with
     $P_{cum}(\tau=\tau_M\Omega)$ [Eq.~(\ref{PcuM})] for the regimes 
    ${\bf C}$ (red-dashed lines) and ${\bf O}$ (blue-dashed-dotted
      lines) for $\delta=10^{-5}$. Different values of $K_2$ are used,
      namely $K_2=0.09$ (regular dynamics) in (a),(e),(i); $K_2=2.6$
      (mixed dynamics) in (b),(f),(j) ; $K_2=4.9$ (mixed
      dynamics) in (c),(g),(k) and $K_2=8.8$ (chaotic dynamics)
      in (d),(h),(l). In all simulations we used $\Omega=100$.} 
  \label{rec}
\end{figure}
\end{widetext}

\subsection{Regimes of ordered and chaotic motion}

Using the classification of regimes it is possible to recognize the dynamics for the 
different decays. This is shown in Fig.~\ref{rec}(e)-(h) for {only} one coupling
strength ($\delta=10^{-5}$), which was chosen because it contains all 
distinct decays we have found. In fact, the scaling properties presented in next 
Section shows that the relevant dynamics is not coupling dependent in the limit of 
small couplings. The chaotic regime ${\bf C}$ is connected to the exponential 
decay for smaller times (not shown in details). The ordered regime astonishing 
reproduces the power-law decays (${\bf O_1}$), plateaus (${\bf O}_2)$, plateaus 
break (${\bf O_4}$), and the asymptotic exponential decay (${\bf O_3,
O_{3^{\prime}}}$),  and the random walk (${\bf O_5}$). Additional simulations 
(not shown) realized for other values of $\delta$, also reveal the connection 
between chaotic regime ${\bf C}$ to the exponential decay for smaller times, and 
the ordered regime ${\bf O}$ to all later decays. 

\subsection{Scaling}
\begin{figure}[!htb]
  \centering
  \includegraphics*[width=0.8\columnwidth]{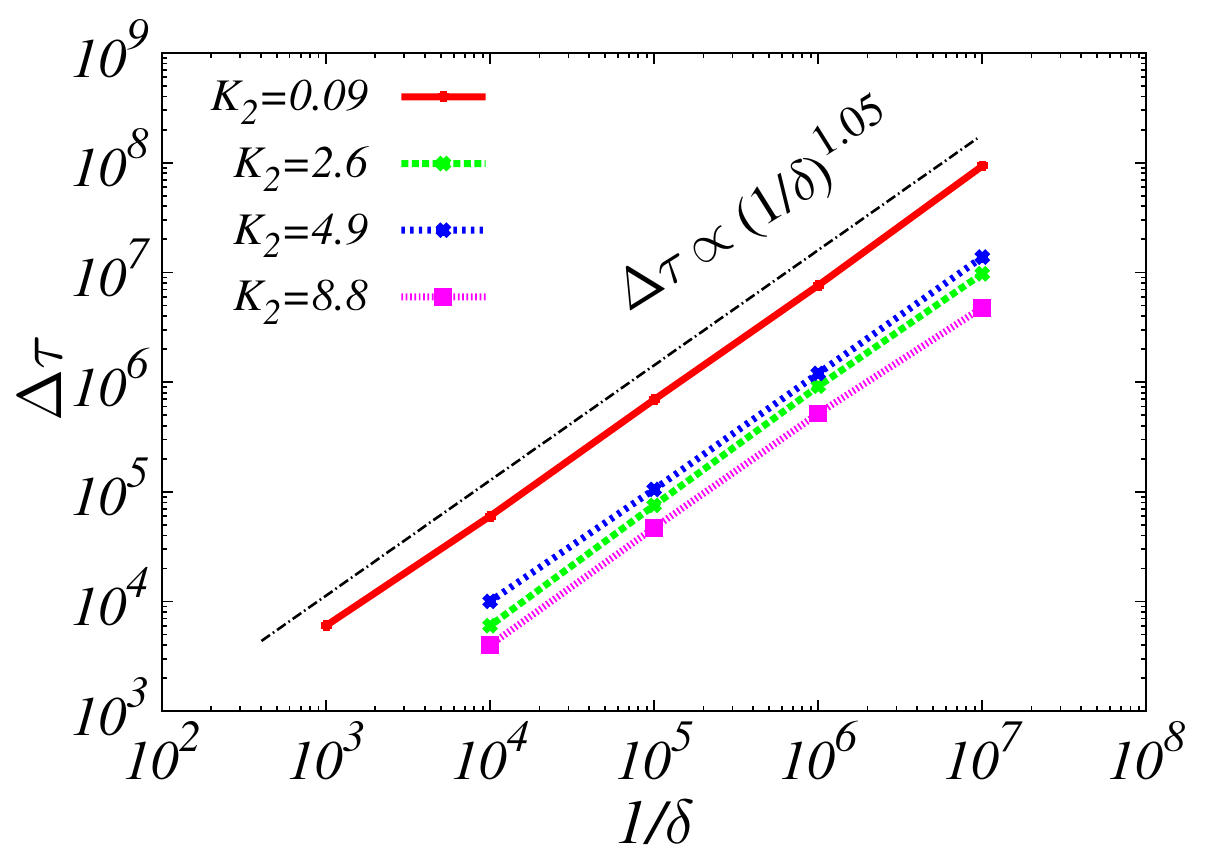}
  \caption{(Color online) Time interval 
  {$\Delta\tau=\tau_{p^{\prime}_{\delta}}-\tau_{p_{\delta}}$}
  where the plateaus exist has a  function of the inverse of the coupling 
  strength. {These times were} obtained from the recurrences curves 
  of Fig.\ref{rec}(a)-(d).}
  \label{delta}
\end{figure}

From the top line of Fig.~\ref{rec} we observe that, for a given $K_2$, curves 
with different $\delta$ are superposed in the exponential (${\bf C}$) and 
power-law decay  (${\bf O_1}$) for short times. Only for intermediate and larger 
times they start to separate from each other. For these times the RTS follows 
the scaling
\begin{equation}
\tau_{\delta} = \frac{\delta^{\prime}}{\delta}\tau_{\delta^{\prime}},\qquad
P_{cum}(\tau_{\delta}) = \frac{\delta}{\delta^{\prime}} P_{cum}(\tau_{\delta^{\prime}}),
\label{scal}
\end{equation} 

\noindent
where $\tau_{\delta}$ ($\tau_{\delta^{\prime}}$) is the
{recurrence-time} for a given coupling strength $\delta$
(${\delta^{\prime}}$). This scaling  
is shown in the bottom line of Fig.~\ref{rec}. The time where the plateaus 
disappear are now exactly the same, independent of the couplings (for small 
couplings). The same scaling occurs in Fig.~\ref{rec}(k), where the abrupt break 
of the plateau and the random walk occurs. This shows that larger couplings 
strongly anticipate the asymptotic decay of the RTS.
Moreover, it is possible to describe the decays of the RTS by the expression

\begin{equation}
P_{cum}(\tau)\sim (a\,\tau^{-\gamma}+b\,\nu)\, e^{-c\,\nu\,\tau},
\label{fit}
\end{equation}

\noindent
where $a,b$ are constants to be determined and $\nu=10^{-2}\delta$. 
Additionally, in Fig.~\ref{delta} we show that the intervals for which the 
plateaus exist $\Delta\,\tau$, has an inverse dependence on $\delta$, 
{\it i.~e.} $\Delta\,\tau\propto (1/\delta)^{1.05}$. This result is almost 
independent of the regular, mixed or chaotic dynamics of the extra-dimension.

\subsection{Diffusion channels and the penetration of islands}

Here we show the dynamics in the $2$D projection in the variables
($x,p$) and the $3$D phase-spaces for specific times, where the 
RTS has distinct decays. This allows us to clearly understand the 
dynamics that occurs at the decays, plateaus, abrupt break of the 
plateaus, random walk and the asymptotic exponential decay. It also 
shows how trajectories use the extra-dimension to penetrate the 
island. 

The $2$D phase-spaces (projections of the dynamics in {$3$D}) are shown in 
Fig.~\ref{ps}, with colors indicating the points plotted during the different 
time decays, as shown in the insets of the figures. The power-law decays for 
times ${\bf O_1}$ are associated to the sticky motion around (but outside) 
the regular structures. The plateaus (${\bf O_2}$) in the RTS are related to 
points in the phase-space where the trajectory remains close to the border (but 
inside) of the structures. This occurs also close to higher-order resonances, 
better visualized in Figs.~\ref{ps}(a),(c) and (d). In the corresponding $3$D 
plots displayed in Fig.~\ref{ps3d}, we observe that for times where the plateaus 
exist, the chaotic trajectory is mainly trapped to the regular structure, but 
deep into the extra-dimension $w$. The trajectory performs a spiralling motion 
along the $w$ direction, leading to the green {trapping tube},  which can be 
better recognized in Figs.~\ref{ps3d}(a),(c) and (d). After the plateaus, 
exponential decays ${\bf O_3}$ occur in Fig.\ref{ps}(a),(b) and (d) which are 
related to the penetration of the trajectory in the regular structures, going 
towards the center. Exception is the penetration of the period-$1$ large structure 
in Fig.~\ref{ps}(d). From Figs.~\ref{ps3d}(a) and (b) we can better see that the 
motions towards the center of the regular structures occur via the extra-dimension, 
repeating the motion along the trapping tube with a smaller radius. 
As mentioned before, the stable period-$1$ and $4$ resonances become 
unstable for $\delta > 0$. Therefore, they act as resonances which eject the 
trajectories along the trapping tube in the extra-dimension, a process which 
remembers the RID. An interesting break of the central structure is observed in the 
$w$ coordinates, due the high-perturbation caused by the extra-dimension, which 
inhibits the penetration of this structure. The additional ${\bf O_3^{\prime}}$ 
exponential behavior occurs only in Fig.~\ref{ps}(b), and is solely related to the 
penetration of the higher-order resonances period-4 islands from the uncoupled case. 
This is also nicely confirmed in Fig.~\ref{ps3d}(b). The short time superdiffusive 
decay ${\bf O_4}$ in Fig.~\ref{ps}(c) is also related to the penetration of the 
regular structures, but the injection via the tube in the extra-dimension is clearly 
distinct, as seen in Fig.~\ref{ps3d}(c), from which we conclude that the trajectory 
first reach the center of the period-$4$ and period-$1$ structures, and after that it is 
ejected along the tubes. 
\begin{widetext}
$\quad$
\begin{figure}[!htb]
  \centering
 \includegraphics*[width=1\columnwidth]{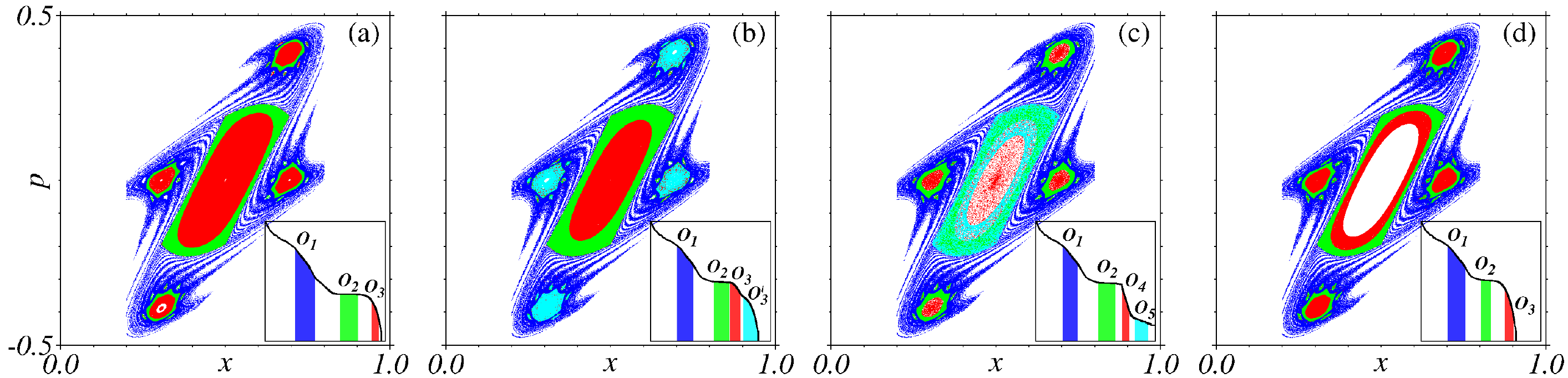}
 \caption{(Color online) Phase-space projected in ($x,p$) for
   $\delta=10^{-5}$ and (a) $K_2=0.09$ (regular dynamics) (b) $K_2=2.6$
    (mixed dynamics) (c) $K_2=4.9$ (mixed dynamics) and (d)
    $K_2=8.8$ (chaotic dynamics). The insets show the RTS, the 
   same insets from the top line of Fig.~\ref{rec}, but now  with
    colors indicating the different time intervals ${\bf O_1}, {\bf O_2},
     {\bf O_3}, {\bf O_3^{\prime}}, {\bf O_4}$ and ${\bf O_5}$, for which the
    corresponding phase-space points were plotted with the same color.}
  \label{ps}
\end{figure}
\end{widetext}

\begin{figure}[!htb]
  \centering
 \includegraphics*[width=0.9\columnwidth]{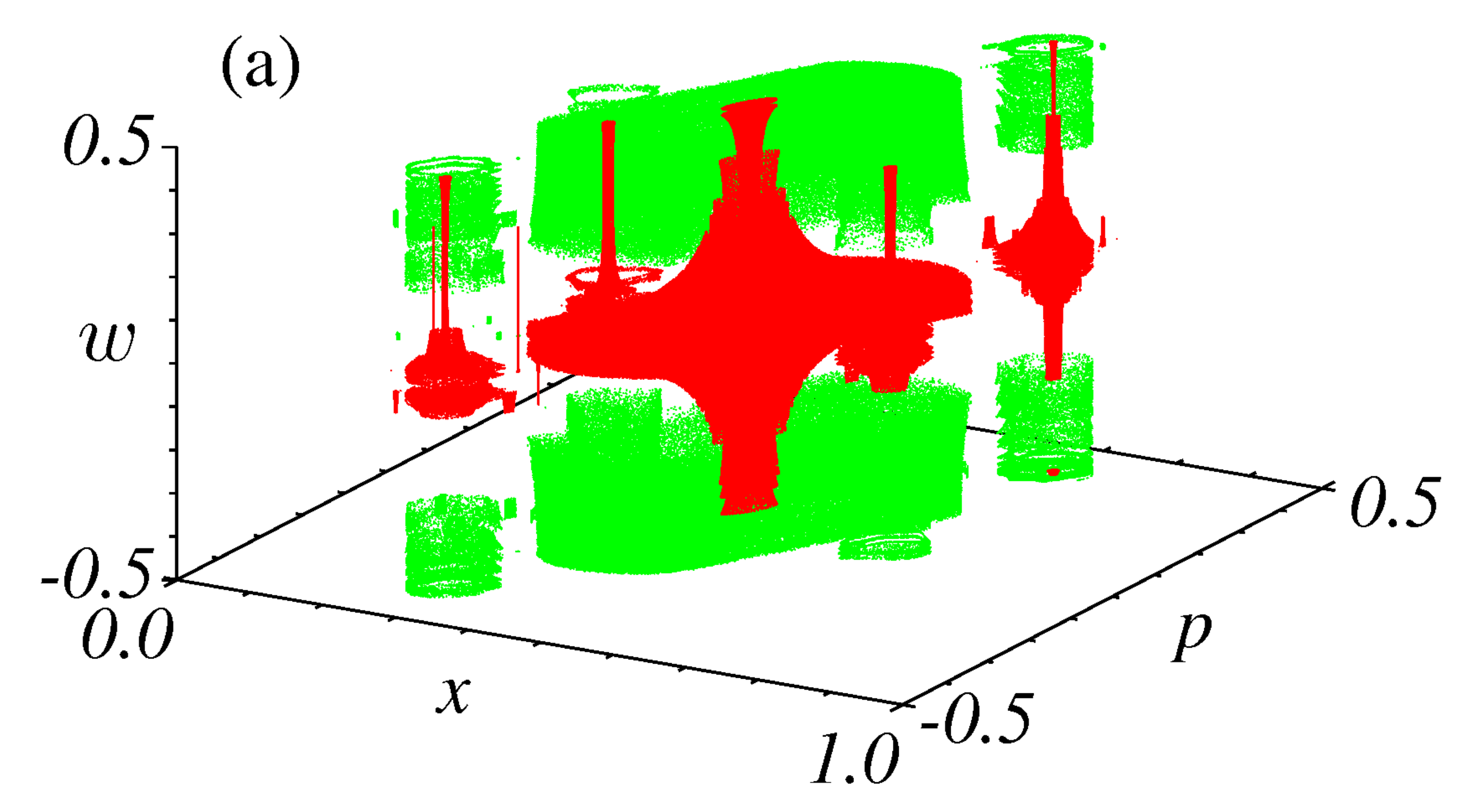}
 \includegraphics*[width=0.9\columnwidth]{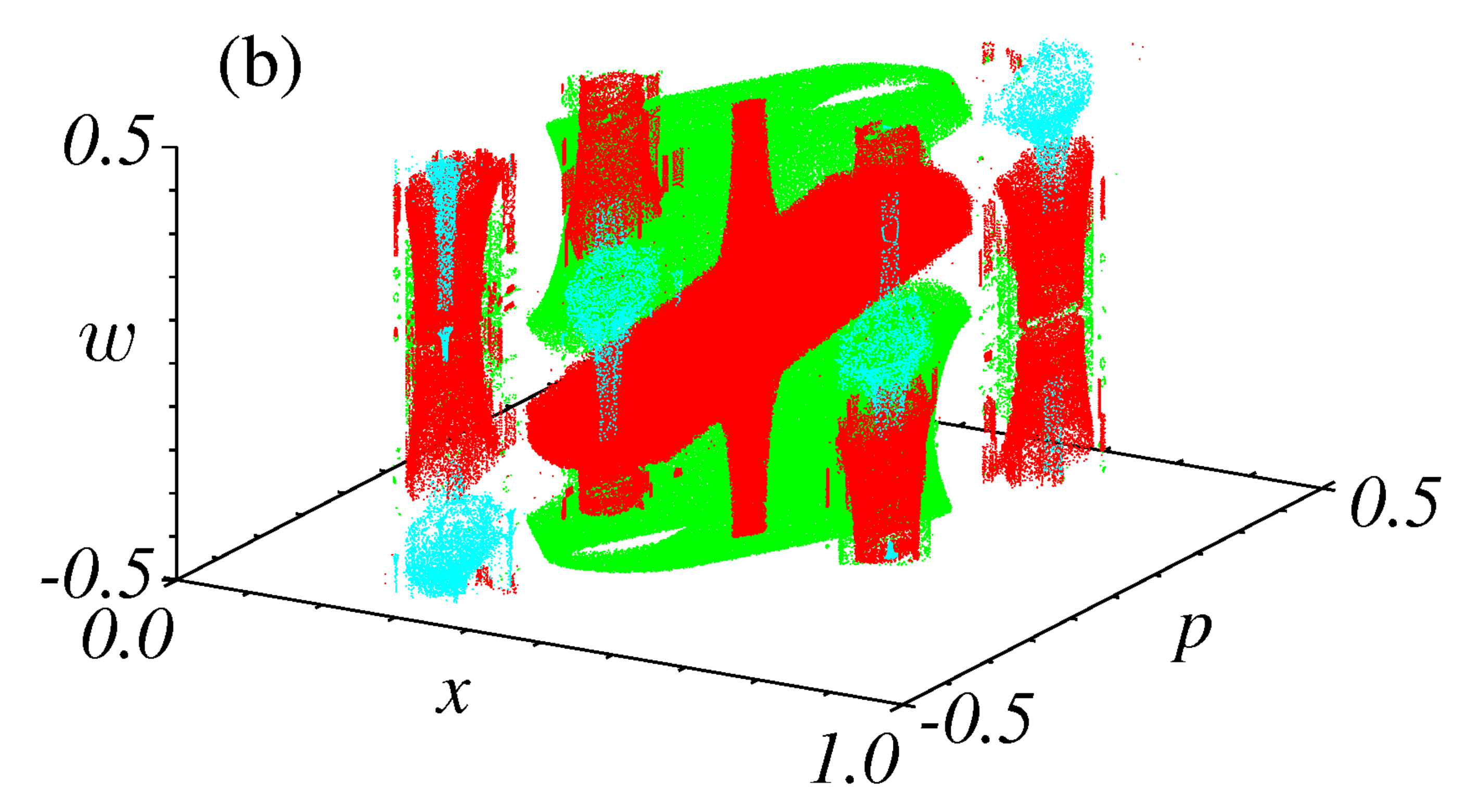}
 \includegraphics*[width=0.9\columnwidth]{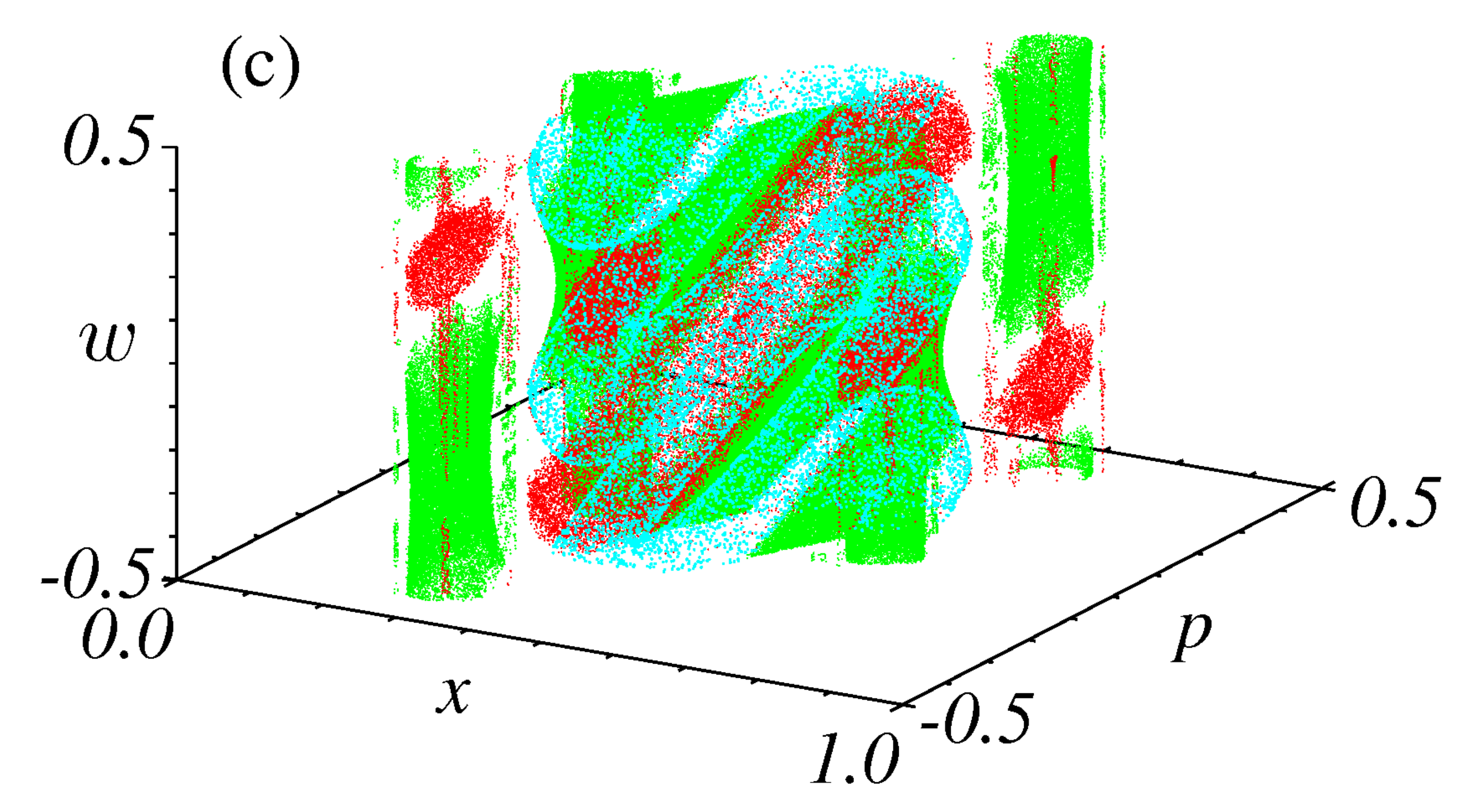}
 \includegraphics*[width=0.9\columnwidth]{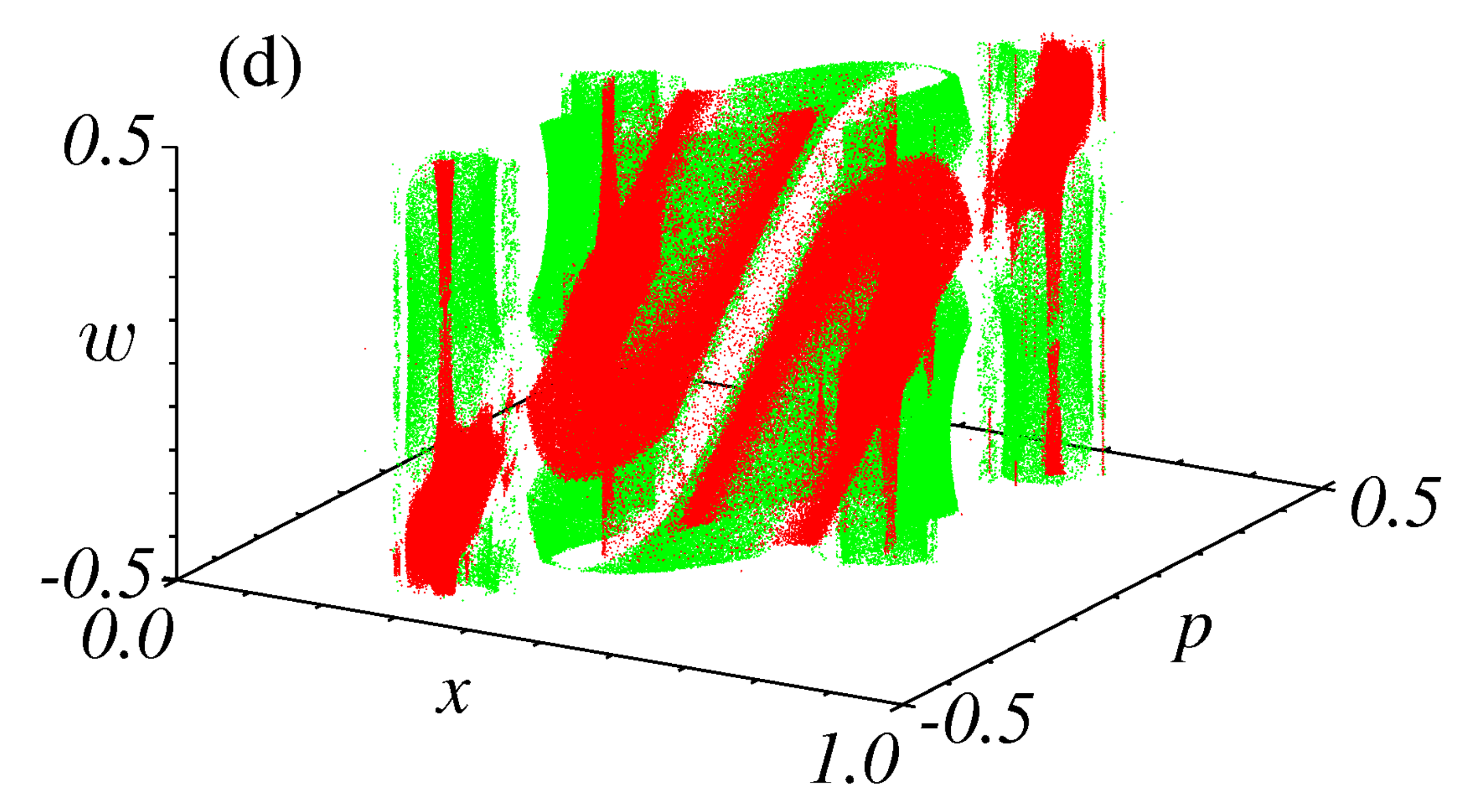}
 \caption{(Color online) Three-dimensional phase-space ($x,p,w$) 
 for (a) $K_2=0.09$ {(regular dynamics)}; (b) $K_2=2.6$ {(mixed
   dynamics)}; (c) $K_2=4.9$ {(mixed dynamics)} and (d) $K_2=8.8$
 {(chaotic dynamics)}. These plots correspond respectively to
 Figs.~\ref{ps} (a)-(d). Color denote the same time intervals from
 Fig.~\ref{ps}. For clarity the times ${\bf O_1}$ (blue points) from  
 Fig.~\ref{ps} were not plotted.}
 \label{ps3d}
\end{figure}

The last observed power-law decay, ${\bf O_5}$, occurs only inside the main
regular structure and is related to a random walk motion restricted to the 
boundaries imposed by the regular structure. 
This occurs just for finite times and we expect an exponential decay 
for larger times (not shown), as nicely discussed in \cite{eduardoPRL10}
for the case of noise.
Amazingly, all the dynamics shown in Figs.~\ref{ps} and \ref{ps3d} are related 
to the approximated regular motion which belongs to the regime ${\bf O}$. 
Even the asymptotic exponential decays of the RTS are related to this 
regime and suggest the mixing nature of the dynamics 
\cite{eduardoPRL10}. 

\section{Uncoupled case with dissipation or noise}
\label{uncoupled}

\begin{figure}[!htb]
  \centering
  \includegraphics*[width=0.99\columnwidth]{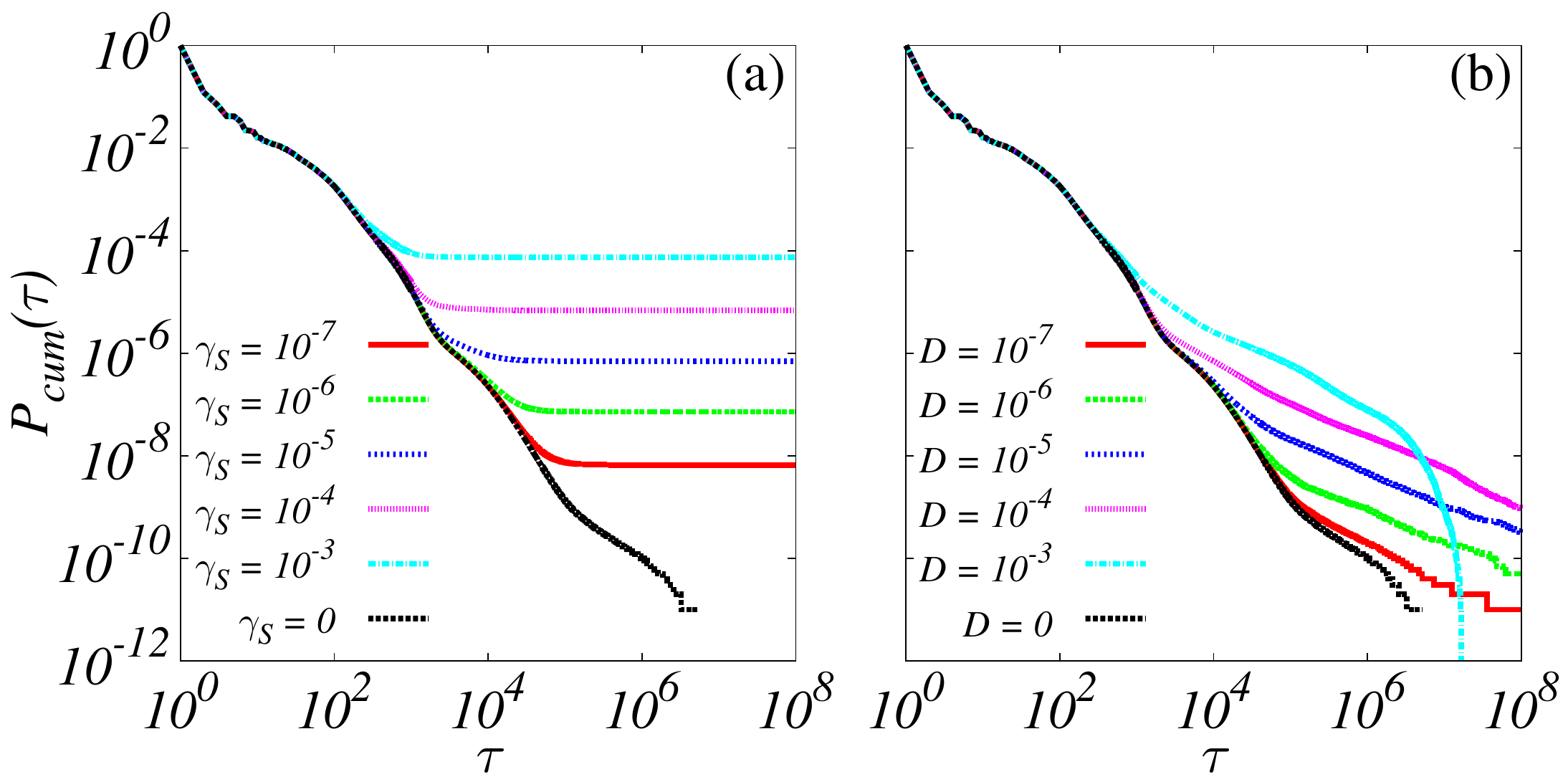}
  \caption{(Color online) {Cumulative probability distributions for
     recurrence-times $\tau$} for distinct values of (a)
    dissipation {$\gamma$}  and (b) noise intensity
    $D$.} 
  \label{diss}
\end{figure}

In this Section we compare our results of a deterministic perturbation in the 
standard map to other kind of perturbations. {It has been argued that
noise and dissipation can play the role of an extra-dimensional
environment. Therefore we are now interested in comparing the RTS from
our extra deterministic dimension with the RTS by using noise or
dissipative extra-dimensions. With this purpose in mind we take the
following} map 
\renewcommand{\arraystretch}{2.2}
\begin{equation}
  \label{dsm}
  \left\{
  \begin{array}{lllllll}
    p_{n+1} & = & \left(1-\dfrac{\gamma_S}{2\pi}\right)\, p_n + 
    \dfrac{K}{2\pi} \sin{(2\pi\,{x_n})} + \dfrac{D}{2\pi} \xi_n, && \\
    x_{n+1} & = & x_n + p_{n+1},
  \end{array}
  \right.
\end{equation}
\renewcommand{\arraystretch}{1}
which is the standard map with dissipation and noise. The dissipation 
constant is $0\le\gamma_S\le1$ and $\xi_n$ is the white noise, equally 
distributed in the interval [$-1,1$], with intensity $D/2\pi$. 
The standard map close to the conservative limit has a larger number of 
resonances ($\sim 1/\gamma_S$ ) which essentially determine the kind of 
dynamics. {All regular trajectories surrounding the resonances will 
converge asymptotically to the center of the stable islands.}
The effect of dissipation on the RTS with $D=0$ is shown in 
Fig.~\ref{diss}(a), demonstrating that dissipation induces plateaus which 
remain for all the iterated times. In other words, the chaotic trajectory 
only returns to the recurrence region for smaller times. For larger times 
it is attracted by the elliptic points from the conservative case which 
were transformed into sinks due to the dissipation. {Thus plateaus 
are related to trajectories which converge to the sinks and never return.} 
The times $\tau_{\gamma_S}$ and $\tau_{\gamma^{\prime}_S}$ for which
the plateaus appear for different dissipation parameters ${\gamma_S}$
and $\gamma^{\prime}_S$ are scaled by  
\begin{equation}
  \tau_{\gamma_S} =
  \frac{\gamma_S}{\gamma_S^{\prime}}{\tau_{\gamma^{\prime}_S}} 
\end{equation}
and are therefore proportional to the number of resonances. This {\it
  confirms} that the plateaus observed in Sec.~\ref{sec:results} are
also related to the penetration of the island. However, in that case
the center of the regular structure are 
unstable points and eject the trajectory back
to the recurrence region via the extra-dimension. Thus, to
return to the recurrence region the trajectory needs to reach the 
center of the structure and, therefore, the plateau times observed
in Sec.~\ref{sec:results} are the times the trajectory need to penetrate
the regular structure. These times are proportional to 
$\sim1/\delta$. 

Figure \ref{diss}(b) displays the RTS for the case of noise with $\gamma_S=0$. 
Small noise intensities induce an enhanced power-law decay which tend to 
disappear for larger values of $D$. For $D=10^{-3}$ we observe that, in addition 
to the power-law decay, an asymptotic exponential decay occurs.  Such asymptotic 
exponential decays were also observed when noise is included in open chaotic 
billiards \cite{altmann12}. The case with $D\ne\gamma\ne0$ (not shown here) 
was already studied recently \cite{klages14} and the RTS mainly follows the 
behavior observed in Fig.~\ref{diss}(b). {Thus, while noise may 
induce the asymptotic exponential decay observed in volume preserving systems, 
sinks generate the plateaus.}

\section{RTS in a fluid flow model}
\label{flow}
For time-periodic $2$D flows and steady $3$D flows, regular and chaotic motions
coexist and the increasing of mixing, resulting from the chaotic advection, is 
forbidden due the impenetrable barriers that separate the two distinct motions. 
However, for time-dependent $3$D flows, complete uniform mixing is possible due 
to the {\it resonance-induced dispersion}, as mentioned in 
Sec.~\ref{sec:introduction}. In this case an enhancement of diffusion in some 
regions of the space will occur \cite{passivescalars}. The previous Sections 
showed the existence of these behaviors in the dynamics of the {\it ESM}, which lead 
us to conjecture that some aspects of the dynamics of a time-dependent $3$D flow 
can be reproduced by a $3$D volume preserving maps of the type
action-action-angle. 
  
In this Section we investigate the dynamics of a time-dependent
non-Hamiltonian $3$D flow model. We perform the RTS simulations 
for this model and compare the results with those described in Sec.~\ref{RTS}. 
The connection between  RTS and transport in fluids is by itself an
interesting problem \cite{zas91}. The model used here is given by
\cite{mezic03} 

\renewcommand{\arraystretch}{1.8}
\begin{equation}
\label{flow-model}
\begin{array}{lllllll}
v_x=-\cos(\pi x_s(t))\sin(\pi y) + {\epsilon} \sin(2\pi
x_s(t))\sin(\pi z), \\

v_y=\sin(\pi x_s(t))\cos(\pi y) + {\epsilon} \sin(2\pi y)\sin(\pi z),\\

v_z=2{\epsilon} \cos(\pi z)[\cos(2\pi x_s(t)) + \cos(2\pi y)],
\end{array}
\end{equation}
where ($x,y,z$) and ($v_x,v_y,v_z$) represent respectively the fluid position
and velocity. The explicit time dependence is given by $x_s(t)=x+b\,\sin(\omega t)$, 
where $b$ and $\omega$ are the non-dimensionalized oscillation amplitude and 
frequency, respectively. The coupling to the {extra-dimension (secondary 
motion of the} flow) is characterized by {$\epsilon$}. This model captures the 
essential features of an alternating vortex flow with Ekman pumping. {For the 
two-dimensional uncoupled case {$\epsilon=0$}, the motion is clearly separated 
in regular (vortices) and chaotic around the vortices. A tracer inside the chaotic 
region never crosses the regular region. However, tiny values of {$\epsilon$} 
are enough to induce a nearly uniform mixing. This mixing occurs via a spiralling 
motion (see \cite{mezic03} for more details). }
\begin{figure}[!htb]
  \centering
  \includegraphics*[width=0.7\columnwidth]{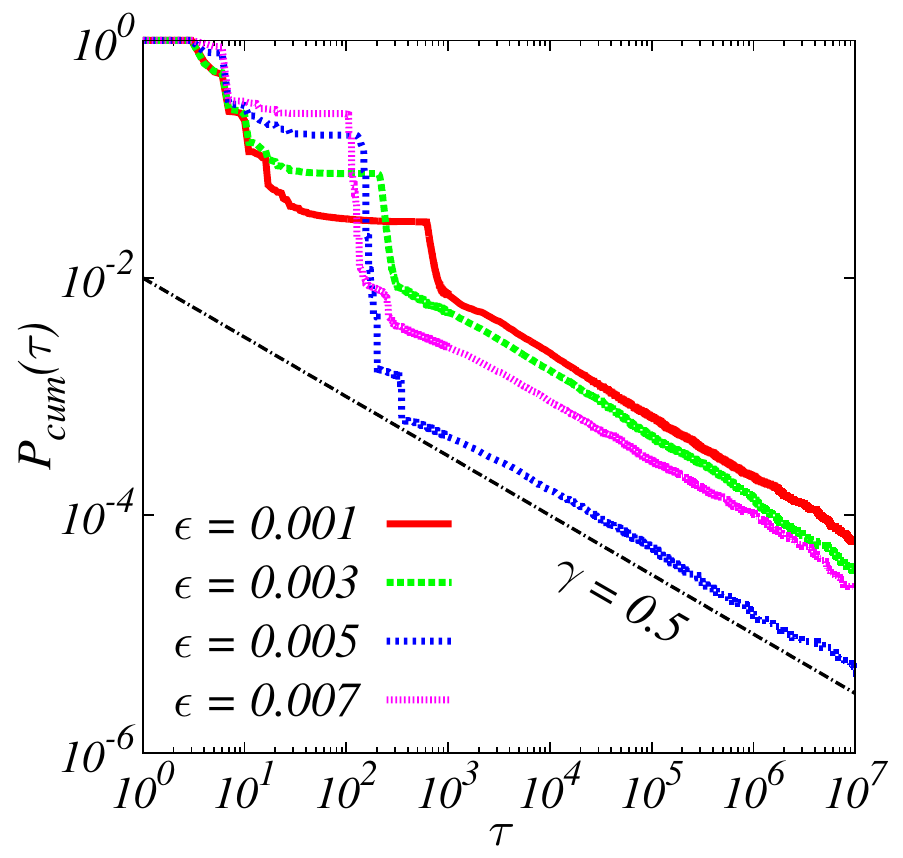}
  \caption{(Color online) {Cumulative probability distributions for
     recurrence-times $\tau$} for the flow described by
    Eq.~(\ref{flow-model}) using $b=0.02$, $\omega=4.0$ and different
    values of coupling parameter $\epsilon$.}
  \label{flow-fig}
\end{figure}

Figure \ref{flow-fig} shows the RTS curves for the model (\ref{flow-model}). In
our simulations we use fourth-order Runge-Kutta algorithm with fixed time-step
$\Delta t=0.01$ for $10^6$ recurrences. {The similarities of these decays with 
Fig.~\ref{rec}(c),(g) and (k) are astonishing. The general sequence of decays 
is identical. Plateaus followed by an abrupt break and a very long random walk 
decay. The existence of plateaus is induced by the spiralling motion (trapping 
tube). After the abrupt break of the plateaus, particles from the fluid performs 
a random walk clearly characterized by the power-law 
$P_{cum}(\tau) \propto \tau^{-0.5}$. Thus the mixing process in the fluid 
dynamics is mainly governed by a random walk process.} Trapping regime for short 
times, originating plateaus on the RTS curves as we found for the {\it ESM} in 
Fig.~\ref{rec}. Additionally, the influence of the parameter $\epsilon$ can be 
compared to the parameter $\delta$ for the {\it ESM}. By increasing the
parameter the plateaus appear early and become shorter.

Another example of this behavior can be found in \cite{Khurana}, as mentioned in 
Sec.~\ref{sec:introduction}. For the model used by the authors, the interaction of 
the active particles with the fluid flow induces a complex behavior which cannot 
be reproduced by the individual systems. The dynamics of the fluid flow alone 
is composed of a chaotic sea and elliptic islands bounded by KAM tori that are 
impenetrable for the fluid elements. However, when it is coupled with the active 
particles dynamics, the swimmers can cross these boundaries. Transport decreases 
due to the formation of traps that can stuck the swimmers on nearly bounded 
orbits for long times \cite{Khurana}, a feature similar to our observation in the 
{\it ESM} dynamics. Furthermore, some of these behaviors of flows and maps 
can also experimentally be observed for passive scalars in the Rayleigh-Benard 
system with oscillatory instability \cite{PiroFeingold}.

\section{Conclusion}
\label{sec:conclusion}

This work uses the RTS to describe the regular, mixed and chaotic dynamics 
in non-Hamiltonian $3$D volume preserving systems. For this we use the 
standard map coupled to an extra-dimension and a continuous fluid flow model. 
The standard map is a typical system which allows us to make important general 
statements about the diffusion process steering the penetration 
of islands from the $2$D case. In our case we do not have Arnold diffusion, but 
a process similar to the resonance-induced dispersion takes 
place \cite{CartwrightFeingoldPiro,msm1}. The general observed behavior for the 
decays of the RTS can be divided in two parts. Firstly, an initial exponential 
decay due to the chaotic regime ${\bf C}$. This is a consequence of trajectories 
which do not touch the regular structure and are of no interest here. 
Second, the dynamics in the  ordered regime ${\bf O}$ (quase-regular motion), 
which is a consequence of particles which interact with the regular structure. 
This part contains power law decays due to sticky effects around the
regular structure, plateaus due to trapping inside the regular structures 
(in this case a trapping tube or vortices in the fluid flow case), and asymptotic 
exponential decays. In addition to the above decays, 
when the mixed dynamics is considered in the extra-dimension, we also observe an 
abrupt break of the plateau due to a diffusive motion, followed by a random walk. 
The trapping times inside the regular structure
are clearly recognized by the plateaus in the RTS and in the cumulative
distribution of consecutive time spend inside the regime ${\bf O}$. The large 
trapping times occur independent of the regular, mixed or chaotic dynamics of 
the extra-dimension and tends to infinity (but with decreasing probability to 
occur) for smaller and smaller couplings. The plateau time is shown 
to be inverse proportional to the coupling strength between the 
standard map and the extra-dimension. Compared to RTS in Hamiltonian 
systems, we observe that for mixed $3$D volume-preserving systems the asymptotic
exponential decays appear naturally (with one exception for $K_2=4.9$, but we believe 
this is a question of iteration times) and that plateaus seem to be a unique property 
of such $3$D non-Hamiltonian conservative systems. Scaling properties show 
that the asymptotic behavior of decays and regimes is independent of the small 
coupling values. We also show that plateaus with infinite times appear when tiny 
dissipation is introduced in the standard map, and are consequences of the 
larger number of sinks. On the other hand, the standard map with a white
noise is shown to induce an enhanced power-law decay followed by an asymptotic 
exponential decays with no plateaus.

The classification technique \cite{marques15} of ordered and chaotic regimes
allows us to show that all relevant dynamics occurs for the ordered regime, 
or quasi-regular dynamics. Asymptotic exponential decays of trajectories 
moving restricted to the boundaries of the islands from the uncoupled
case suggest that the dynamics in this system belongs to the mixing one. 
Further interesting investigations is to use an extra-dimension 
for which the stable points remain stable. In such case the diffusion 
process is expected to differ from the observed here and it may change
the asymptotic decay.

\section*{Acknowledgments}
The authors thank FINEP (under project CTINFRA-1), C.M. and M.W.B. thank CNPq for 
financial support and M.W.B. thanks UDESC for the hospitality in November, 2014.


\end{document}